\documentclass[doublecol]{epl2} 

\usepackage{amsfonts}
\usepackage{mathrsfs}
\usepackage{amsmath}
\usepackage{geometry}
\usepackage{graphicx}
\usepackage{verbatim}
\usepackage{appendix}

\title{Eco-evolutionary dynamics with environmental feedback: cooperation in a changing world}
\shorttitle{eco-evolutionary dynamics with environmental feedback} 

\author{Xin Wang\inst{1,2,3}\thanks{E-mail: \email{Xin.Wang-2@dartmouth.edu}}\and Feng Fu\inst{3,4}\thanks{E-mail: \email{feng.fu@dartmouth.edu}} }
\shortauthor{X. Wang \etal}

\institute{                    
  \inst{1} LMIB, SKLSDE, BDBC and School of Mathematical Sciences, Beihang University, Beijing 100191, China\\
  \inst{2} PengCheng Laboratory, Shenzhen, 518055, China\\
  \inst{3} Department of Mathematics, Dartmouth College, Hanover, NH 03755, USA\\
  \inst{4} Department of Biomedical Data Science, Geisel School of Medicine at Dartmouth, Lebanon, NH 03756, USA
}


\abstract{Eco-evolutionary game dynamics which characterizes the mutual interactions and the coupled evolutions of strategies and environments has been of growing interests in very recent years. Since such feedback loops widely exist in a range of coevolutionary systems, such as microbial systems, social-ecological system and psychological–economic system, recent modeling frameworks that unveil the oscillating dynamics of social dilemmas have great potential for practical applications. In this perspective article, we overview the latest progress of evolutionary game theory in this direction. We describe both mathematical methods and interdisciplinary applications across different fields. The ideas worthy of further consideration are discussed in prospects, with the central role of promoting cooperations in a changing world. }

\begin{document}

\maketitle


\section{Introduction}
Cooperation is the cornerstone of human civilization and is important for the efficient and stable development of economy and society \cite{nash1950bargaining,nowak2006evolutionary}. However, the ``selfish gene" is widespread in all levels of complex life systems which often drives the emergence of social dilemmas in the context of natural selections \cite{riley2002bacteriocins,west2006social,kollock1998social}. Therefore, establishing persistent cooperations has always been the central quest of evolutionary game theory \cite{nowak2004emergence,akcay2018collapse,hu2020rewarding}. For example, many different mechanisms were proposed to deal with the well-known Prisoner's Dilemma in two-player games, such as incorporating the spatial structure, the complex network topology, punishment and award, direct reciprocity, indirect reciprocity and etc ~\cite{fu2008reputation,fu2009partner,hauert2004spatial,maciejewski2014evolutionary,szolnoki2020gradual}. Furthermore, a large amount of works address the ``tragedy of the commons" arising in public goods game with well-mixed population, which greatly promotes the understanding of sustaining group cooperations in the real world \cite{hardin1968tragedy,feeny1990tragedy,chen2015competition,szolnoki2010reward,santos2008social,hauert2006evolutionary,gomez2011evolutionary,rong2009effect,ginsberg2019evolution,wu2014social,wakano2009spatial,liu2018evolutionary}.

While early studies mainly focus on the evolutionary properties of replicator dynamics, the intrinsic evolution process of the ecological environment is paid little attention \cite{schoener2011newest,post2009eco,turcotte2011impact}. Therefore, the eco-evolutionary games that take into consideration the influence of time-dependent environmental changes have attracted great attention in recent years \cite{raymond2012dynamics,szolnoki2014coevolutionary,stewart2014collapse,li2016coevolution,szolnoki2018environmental}. Since this paper is surely unable to cover everything, we suggest a mini-review of such coevolutionary models \cite{perc2010coevolutionary}.

Recently, a novel theoretical framework that further describes the complex interactions between strategies and the environment is proposed \cite{weitz2016oscillating,tilman2020evolutionary,shao2019evolutionary,wang2020steering}. The core idea is to study the effect of strategy-dependent environmental feedbacks towards resolving social dilemmas, which is based on the fact that the cooperation behavior has a non-negligible power on reshaping the environment while in contrary the environment change also impacts individual decisions of cooperation. This framework successfully explains the oscillating dynamics of both population cooperations and environmental states. 

In this perspective article, we aim to provide an overview of this newly outlined strategy-dependent feedback-evolving games. We will discuss the recent advances in this filed, both from the views of modeling methods and the real-world applications. And finally we will provide an explicit prospect for future studies, including a highlighted open problem of controlling the co-evolutions which could help defeat a series of social dilemmas \cite{wang2020steering}.
 
\section{Modeling frameworks}

\textit{Feedback-evolving dynamics in two-player games.}

\begin{figure}[!h]
    \centering
    \includegraphics[width=0.49\textwidth]{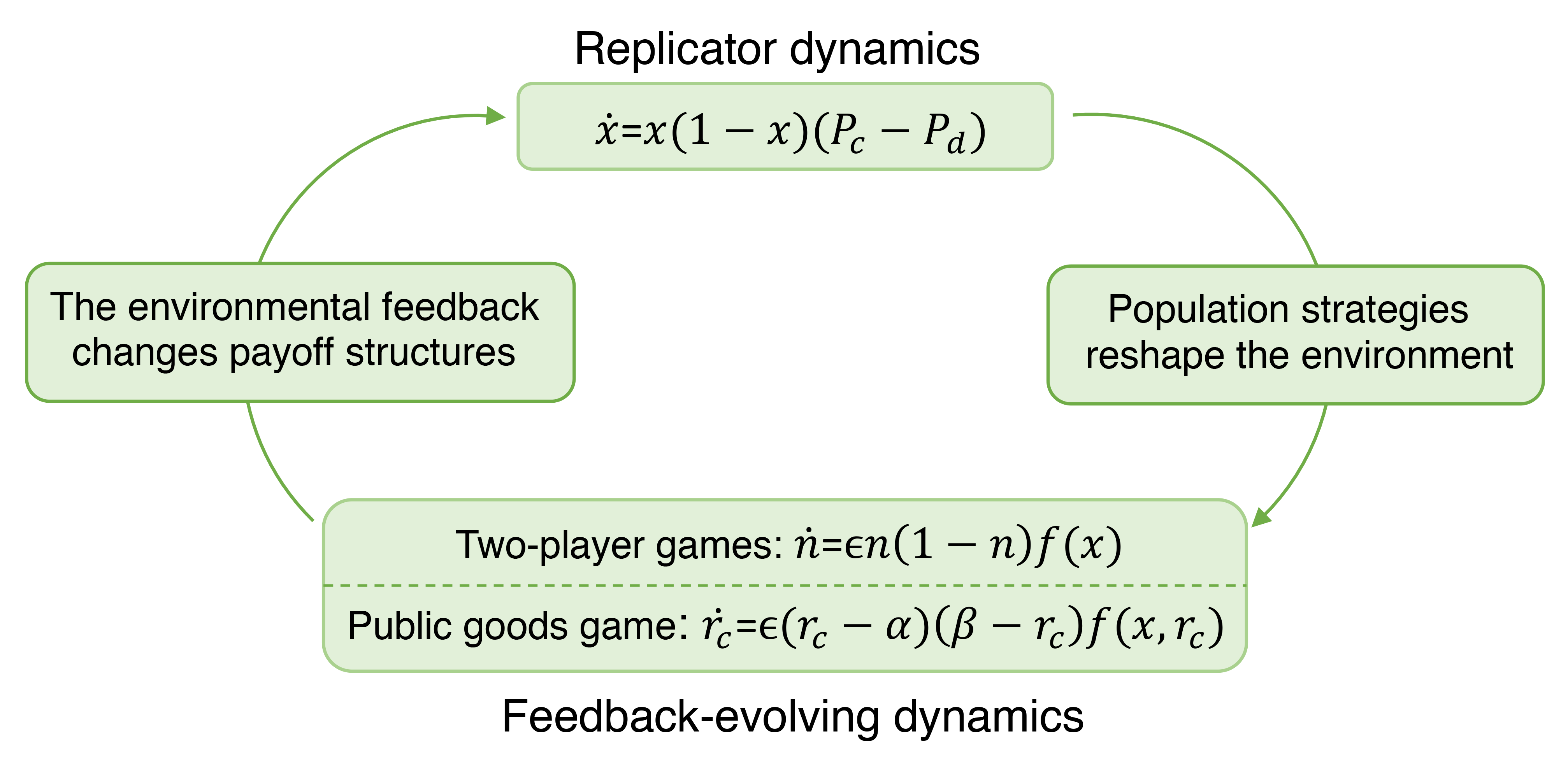}
    \caption{Schematic of evolutionary game theory with environmental feedback. The replicator dynamics determines cooperation behaviors, which reshapes the ecological environment in two-player games (via environment factor $n$) \cite{weitz2016oscillating,tilman2020evolutionary} or the resource allocation in public goods game (via multiplication factor $r_c$) \cite{shao2019evolutionary,wang2020steering}. In turn, the strategy-dependent environmental feedbacks influences the payoff structures of cooperators and defectors and further drives the oscillation of social dilemma. The timescale $\epsilon$ is a nontrivial parameter that describes the relative speed of environmental feedbacks.}
    \label{schematic}
\end{figure}

A unified approach to understand the feedback-evolving games in which the strategies and the environment coevolve is proposed by Weitz \textit{et al}. in \cite{weitz2016oscillating}. We begin from the introduction of a generalized environment-dependent payoff structure in two-player games:
\begin{equation}
	\begin{split}
		A(n)&= (1-n) A_0 +n A_1\\
		       &= (1-n) {\left[\begin{array}{cc}
                                      R_0 & S_0\\
                                      T_0 & P_0
                                    \end{array} 
                                     \right]}  +  n {\left[\begin{array}{cc}
                                      R_1 & S_1\\
                                      T_1 & P_1
                                    \end{array} 
                                     \right]} \\
        \end{split}
	\label{PNAS_e1}
\end{equation}
where $0\leq n \leq 1$ denotes the current state of the environment. $A_0$ and $A_1$ represent the payoff matrices that have a unique Nash equilibrium corresponding to mutual cooperation and mutual defection, respectively. Therefore, $R_0>T_0$ and $S_0>P_0$ while $R_1<T_1$ and $S_1<P_1$. Intuitively, when $n=0$, i.e., the environment is depleted, incentives are offered to encourage cooperation behaviors so that the resources can be restored. On the other hand, when $n=1$, i.e., the environment is replete, the payoffs favor unilateral defection, which accounts for the overuse or overexploitation of a resource. This environment-dependent payoff matrix can be further written as 
\begin{equation}
	\begin{split}
		A(n) = {\left[\begin{array}{cc}
                                      R_0-(R_0-R_1)n & S_0-(S_0-S_1)n\\
                                      T_0-(T_0-T_1)n & P_0-(P_0-P_1)n
                                    \end{array} 
                                     \right]} 
        \end{split}
	\label{PNAS_e2}
\end{equation}
Thus, the fitnesses of player $1$ and player $2$, denoted as $r_1$ and $r_2$, can be calculated as 
\begin{equation}
	\begin{cases}
	\begin{split}
		r_1(x,A(n)) &= x(R_0-(R_0-R_1)n)\\
		                  &+ (1-x)(S_0-(S_0-S_1)n)\\
		r_2(x,A(n)) &= x(T_0-(T_0-T_1)n)\\
		                  &+ (1-x)(P_0-(P_0-P_1)n)\\
	\end{split}
	\end{cases}
	\label{PNAS_e3}
\end{equation}
Finally, the replicator dynamics of this coupled evolutionary system are described as follows
\begin{equation}
	\begin{cases}
	\begin{split}
		\epsilon \dot{x} &= x(1-x)\left[ r_1(x,A(n))-r_2(x,A(n))\right]\\
		\dot{n} &= n(1-n)f(x)\\
	\end{split}
	\end{cases}
	\label{PNAS_e4}
\end{equation}
where $\epsilon$ represents the relative speed that individual actions reshape the environment. The logistic term $n(1-n)$ guarantees that the environment state is confined to $[0,1]$. The sign of feedback function $f(x)$ determines the evolution direction of $n$, either decreases towards environmental degradation or increases towards environmental enhancement when $f<0$ or $f>0$, respectively. In \cite{weitz2016oscillating}, the environment state is assumed to be modified by the population actions of cooperation and a simple linear feedback mechanism is adopted:
\begin{equation}
	\begin{split}
		f(x) = \theta x-(1-x)
	\end{split}
	\label{PNAS_e5}
\end{equation}
where $\theta>0$ indicates the strength of cooperators in enhancing the environment. See the complete modeling schematic in fig. \ref{schematic}.

Under this framework, Weitz \textit{et al.} show the emergence of persistent oscillating loops in a special case where the payoff matrices $A_0$ and $A_1$ have an embedded symmetry. The orientation of orbits in $x-n$ phase plane is always counterclockwise, which in intuitive sense can be explained as four phases that the system evolution experiences: beginning from a depleted environment with few cooperators, to motivated cooperation, to a restored environment, to invading defection, and finally going back to the depleted environmental state. Further, they extend the analysis to asymmetric payoffs conditions and find that the fast-slow ($0<\epsilon \ll 1$) dynamics can converge both to a heteroclinic cycle and to a fixed point. The former is called ``oscillating tragedy of the commons" and it emerges when
\begin{equation}
	\begin{split}
		\frac{P_1-S_1}{T_1-R_1}>\frac{S_0-P_0}{R_0-T_0}
	\end{split}
	\label{PNAS_e6}
\end{equation}

It is also worth noting that the qualitative outcome of coupled systems in \cite{weitz2016oscillating} is independent of the relative feedback speed $\epsilon$. However, in contrary, Tilman \textit{et al.} proposed a more general framework of eco-evolutionary games and found that the dynamical behaviors actually largely depend on the relative timescale \cite{tilman2020evolutionary}.  Three different environmental dynamics are incorporated in their work: intrinsic resource growth, intrinsic resource decay, or environmental tipping points. Generally, this eco-evolutionary system is written as 
\begin{equation}
	\begin{cases}
	\begin{split}
		\dot{x} &= \epsilon_3 x(1-x)\left[ \pi_1(x,n)-\pi_2(x,n)\right]\\
		\dot{n} &= \epsilon_1 f(n) + \epsilon_2 h(x,n)
	\end{split}
	\end{cases}
	\label{NC_e1}
\end{equation}
where $\epsilon_1$, $\epsilon_2$ and $\epsilon_3$ represents the timescales of intrinsic dynamics of the environment, the timescale of the environmental impact of the current population strategies and the timescale of strategy evolution, respectively. In particular, $f(n)$ governs the intrinsic dynamics of the environmental factor $n$ while $h(x,n)$ aggregates the influence of population cooperation behaviors on environmental changes.  And $\pi_1$ and $\pi_2$ are the fitnesses of cooperation and defection, respectively. Accordingly, a renewable resource can be described as
\begin{equation}
		\dot{n} = \epsilon (r-q(e_L n+e_H (1-n))) (x-n)
	\label{NC_e2}
\end{equation}
where $\epsilon$ is the unified relative timescale, $r$ is the intrinsic resource-renewing rate. $e_L$ and $e_H$ are the resource harvest effort of low- and high-effort strategies, respectively. And $q$ maps harvesting efforts into the rate of resource reduction. See \cite{tilman2020evolutionary} for detailed derivation. Similarly, a decaying resource can be calculated as
\begin{equation}
		\dot{n} = \epsilon \alpha (x-n)
	\label{NC_e3}
\end{equation}
here $\alpha$ is the decay rate of the resource. Finally, the environment feedback described in \cite{weitz2016oscillating} can be considered as a special case caused by a single environmental tipping point without any intrinsic environmental dynamics. Based on this framework, Tilman \textit{et al.} show that all the abundant physical phenomena arose from the feedback-evolving system can be interpreted by the incentives for individual behavioral changes and the relative timescale of environmental versus strategic changes. 
 
The modeling extensions that takes into consideration the effects of spatially structured interactions or the environmental heterogeneity can be further obtained in \cite{lin2019spatial} and \cite{hauert2019asymmetric}, respectively.

\textit{Coevolutionary dynamics in public goods game.}

It has been experimentally proven that eco-evolutionary feedback loops widely exist in microbial systems where cooperations often arise due to the secretion or the release of public goods \cite{cortez2020destabilizing,gore2009snowdrift,kummerli2010molecular,sanchez2013feedback}. In specific, cooperators are naturally given preferential access to these common goods, which leads to the asymmetrical payoff-dependent feedback that drives the coevolution of ecological properties and the strategies. Likewise, such group cooperation can also be obtained in many psychological–economic systems where coevolutionary dynamics is engineered by the asymmetrical environmental feedback. One typical example is the crowdsourcing that aims at completing a project by soliciting contributions from a number of individuals or online communities, in which the focal organizer leads the game and often encourages cooperation via providing a higher payoff structure for cooperators, e.g. promising a higher multiplication factor. While previous works exclusively focus on two-player games, in \cite{shao2019evolutionary}, we propose an extended model where individual strategies coevolve with the multiplication factors in public goods game (PGG) to study the emergence of group cooperation in coevolutionary dynamics (fig. \ref{schematic}). The replicator equation for the fraction of cooperators $x$ is
\begin{equation}
	\begin{split}
		\dot{x} &= x(1-x)(P_c-P_d)\\
		&= x(1-x)\left(-c+\frac{(sx+1)c}{s+1}r_c-\frac{sxc}{s+1}r_d\right)
	\end{split}
\end{equation}
where the focal individual randomly chooses $s$ other participants in a well-mixed infinitely population to join the PGG and  $P_c$, $P_d$ are the expected payoffs of cooperators and defectors. In classic PGG, the cooperators first pay $c$ cost to the common pool, then the total contribution will be amplified by a multiplication factor $r$ and finally equally distributes to every player. For simplicity and without loss of generality, in \cite{shao2019evolutionary} we set the initial contribution of each cooperator be $1$, i.e., $c=1$. In addition, to mimic the asymmetrical feedback-evolving characters, we assume the multiplication factor of defectors $r_d$ is invariant while the multiplication factor of cooperators $r_c$ co-evolves in response to the global payoffs distributions, which is described as
\begin{equation}
	\dot{r_c} = \epsilon (r_c-\alpha)(\beta-r_c) f(x,r_c)
\end{equation}
here we confine $r_c$ in $[\alpha, \beta]$ and $1<\alpha<\beta<s+1$ according to the social dilemma in PGG. $\epsilon$ is the relative feedback speed of $r_c$ versus $x$. Moreover, to describe the fact that in crowdsourcing project the authoritative organizer may decide the global incentives distributions for cooperators and defectors in oder to facilitate the collaboration, we define the feedback mechanism $f(x,r_c)$ as a linear function of global payoffs of cooperators ($xP_c$) and defectors ($(1-x)P_d$) and take into consideration the limitation of total rewards for the project:
\begin{equation}
	f(x,r_c) = -xP_c+\theta(1-x)P_d
\end{equation}
where $\theta>0$ is the distribution ratio of cooperator's and defector's total payoff expectations. Under this circumstance, cooperators are favored through increasing $r_c$ when $x$ is small, whereas the continuous consumption of resources results in the decrease of cooperator's rewards, which is also in line with the law of diminishing marginal utility in economics. The final ODEs for our model can thus be written as
\begin{equation}
	\begin{cases}
	\begin{split}
		\dot{x} &= x(1-x)\left(\frac{sx+1}{s+1}r_c-\frac{sx}{s+1}r_d-1\right)\\
		\dot{r_c} &= \epsilon(r_c-\alpha)(\beta-r_c)\left[-x\left(-1+\frac{r_c(1+sx)}{s+1}\right)\right.\\
		&\left.+\theta(1-x)\frac{r_dsx}{s+1}\right]
	\end{split}
	\end{cases}
	\label{EPL_e1}
\end{equation}

We highlight our main finding that the coevolutionary dynamics with asymmetrical environmental feedback can give rise to oscillating convergence to persistent group cooperation, but only if the feedback updates quickly and promptly enough compared to the strategy change. Mathematically, the unique interior fixed point 
\begin{equation}
	\begin{cases}
	\begin{split}
		x^*&=\frac{\theta}{1+\theta}\\
		r_c^*&=\frac{\theta r_ds+(s+1)(\theta+1)}{\theta s+\theta+1}
	\end{split}
	\end{cases}
	\label{EPL_e2}
\end{equation}
is stable only when the relative feedback speed of cooperator’s multiplication factor exceeds a threshold, i.e., $\epsilon>\epsilon^*$, where $ \epsilon^*$ depends on $s$, $r_d$ and $\theta$:
\begin{equation}
	\epsilon^*=\frac{(1-x^*)s(r_c^*-r_d)}{(sx^*+1)(r_c^*-\alpha)(\beta-r_c^*)}
	\label{epsilon_thredshold}
\end{equation} 
Our work sheds lights on how to successfully organize and sustain a desired group collaboration, which provides useful insights into avoiding the traps of social dilemma in many real scenarios such as knowledge discovery and management, crisis mapping, crowdfunding, scientific cooperation and etc. 

Following the similar idea, a further exploration of coevolutionary dynamics with asymmetrical feedback mechanism in spatial threshold PGG is performed in \cite{wang2020evolutionary}, where simulations are conducted on both square lattice and scale-free networks. The evolutionary dynamics of a nonlinear PGG, which incorporates discounted or synergistically enhanced value of accumulated cooperative benefits \cite{hauert2006synergy}, with different types of ecological variations is also studied in \cite{gokhale2016eco}. Note that the environmental feedbacks in \cite{gokhale2016eco} are all time-dependent, not strategy- or payoff-dependent.

\textit{An open problem: controlling eco-evolutionary games with external feedback laws.}

 \begin{figure*}[!h]
    \centering
     \includegraphics[width=0.95\textwidth]{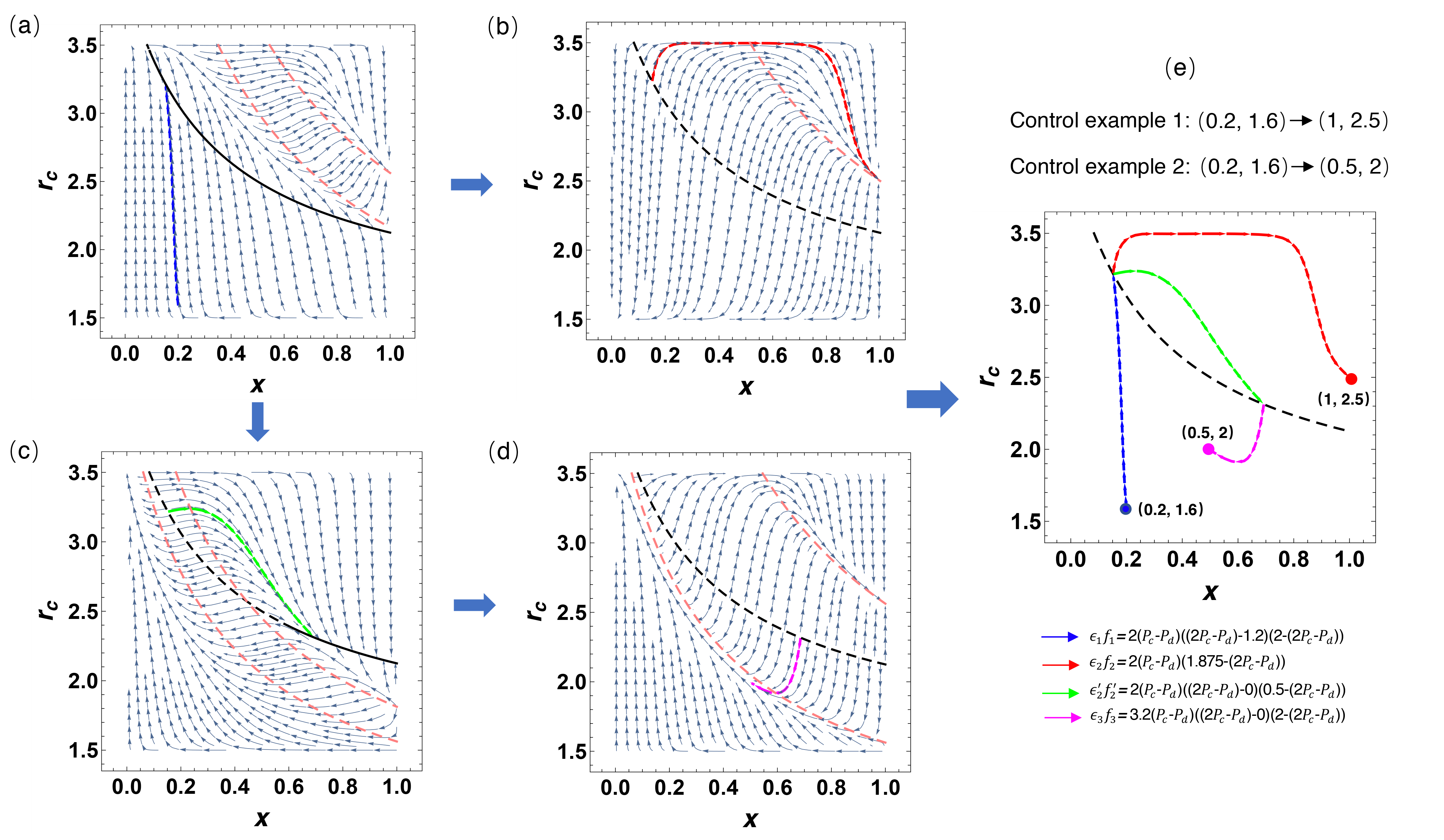}
    \caption{Manifold control on eco-evolutionary game dynamics with external switching control laws. Two detailed examples are presented according to \cite{wang2020steering}. We begin from $(x_0, r_0)=(0.2,1.6)$ and give two desired final state: $(x_1, r_1)=(1, 2.5)$ and $(0.5, 2)$. (a)(b) Manifold controlling process for the first example. (a)(c)(d) Manifold controlling process for the second example. The stable and unstable parts of the equilibrium curves are indicated by solid and dashed black lines, respectively. The external control curves $\Phi_i(x, r_c)=0$ are presented by dashed orange lines. Finally, the piecewise controlled trajectories highlighted by blue, red, green and pink lines are provided in $x-r_c$ phase planes. Taken together, the complete control paths and the corresponding control functions are shown in (e). Parameters: $\alpha=1.5, \beta=3.5, S=3, r_d=1.5$. }
    \label{control}
\end{figure*}

While great efforts have been made to reveal the rich dynamical outcomes of eco-evolutionary games as well as the detailed conditions for breaking the ``tragedy of the commons", a fundamental theoretical gap in how to effectively steer such coevolutionary dynamics to a desired population state with external control laws remain unsolved. In \cite{wang2020steering}, we 
develop a novel and bold manifold control method based on a generalized framework of coevolutionary multi-player games with asymmetrical feedback driven by a nonlinear selection gradient \cite{constable2013stochastic,constable2016demographic}. The general form of our feedback control laws $f(x, r_c)$ is given by 
\begin{equation}
\begin{split}
	f(x, r_c) = &\Phi_0(x, r_c)(\Phi_1(x, r_c)-a_1) \\
	               &*(\Phi_2(x, r_c)-a_2)...(a_n-\Phi_n(x, r_c)),
 \end{split}
\label{e_f}
\end{equation}
in which
\begin{equation}
\begin{cases}
	\begin{split}
		\Phi_0(x, r_c) &= P_c-P_d\\
		\Phi_i(x, r_c) &= \theta_i P_c-P_d
	\end{split}
	\end{cases}
\end{equation}
and $a_i\ge 0$, $\theta_i>0$, $n+1$ denotes the order of the control law. Here $\Phi_0(x, r_c)= P_c-P_d$ is a simplest form of linear selection gradient. Hence Eq. \ref{e_f} represents a sequence of control functions driven by nonlinear selection gradient in a general polynomial form. When $n=1$, we reach to the simplest situation where $f$ is a quadratic function and the eco-evolutionary model reads
\begin{equation}
	\begin{cases}
	\begin{split}
		\dot{x} &= x(1-x)\left(P_c-P_d\right)\\
		\dot{r_c} &= \epsilon (r_c-\alpha)(\beta-r_c)(P_c-P_d)(a_1-(\theta_1 P_c-P_d))
	\end{split}
	\end{cases}
	\label{e1}
\end{equation}

Surprisingly, we find the emergence of multiple segments of stable and unstable equilibrium manifolds in phase graphs with different feedback control functions, which naturally extends the concept of population equilibrium points in previous models to a manifold (i.e. curve) of stable equilibria. In particular, our result of unstable equilibrium manifold circumstance is consistent with the separatrix phenomenon obtained by experimental study in \cite{sanchez2013feedback}. In addition, we find that a larger
relative feedback speed ($\epsilon$) can not only accelerate the convergence process, but also increase the attraction basin of the stable manifolds.

Based on this framework, we further prove the existence of external switching control laws, either time-dependent or state-dependent, for steering the eco-evolutionary dynamics to any desired region when given an initial population state. For a clear understanding, two detailed control examples are provided in fig. \ref{control}. In the first case (fig. \ref{control}(a)(b) and (e)), the controlled evolution path is the blue trajectory followed by the red one. Beginning from $(x_0, r_0)=(0.2,1.6)$, the trajectory first converges to the stable equilibrium manifold with control law $\epsilon_1f_1$. We then change the external control law to $\epsilon_2 f_2$ so that the equilibrium manifold becomes unstable. Finally, a small disturbance is applied and the trajectory will automatically evolve to the final desired state $(x_0, r_0)=(1,2.5)$. Similar steering process is performed on the second case (fig. \ref{control}(a)(c)(d) and (e)), where the evolution path consists of three segments: the blue, green and pink trajectories. See more details in \cite{wang2020steering}.

According to the experiments in \cite{gore2009snowdrift}, $r_c$ in our model can be modulated by changing the histidine concentration in the growth medium, which can manipulate the relative growth rate of the cooperators compared to the defectors. Therefore, our framework can be applied to many microbial experiments, which is of great significance in systems biology and microbial ecology.

\section{Applications}
With the ubiquity of cooperation-environment co-evolutions in natural systems on different scales, coevolutionary games dynamics has important and wide-ranged applications in interdisciplinary fields. Here we just take a glimpse at the significant potential of this ODEs framework towards understanding a series of widely concerned problems: 

(i) \textit{Overexploitation of renewable resources.} In \cite{chen2018punishment}, a feedback-evolving game within which the renewable resource dynamics also depends on the population strategies is proposed to stress the feasibility of maintaining a healthy shape of common-pool resources. The analytical and numerical evidences reveal that before taking control measures such as punishment to solve the overexploitation problem, we should first pay attention to intrinsic growing capacity of the resources. To be specific, a common-pool resource with very limited growth ability cannot recover from a depleted situation even with the usage of tough punishment that can give rise to the population cooperation.

(ii) \textit{Cooperations during a pandemic.} The unprecedented outbreak of SARS-CoV-2 pandemic has caused great loss of human life and the economic society. Moreover, the supportive level that individuals are willing to take behavioral actions (i.e., population cooperation) is of vital importance for the final outcome of disease spreading \cite{fu2011imitation,chen2019imperfect,xingru2020effectiveness}. Conversely, the prevalence of the virus also has strong impact on individual decisions. This co-evolving loop can naturally be described as an eco-evolutionary dynamics with disease-based environmental feedback. Some inspirational works that incorporate disease transmission models (SIS, SEIR) into cooperation evolution and stress the effects of interactions between disease spreading and social dynamics can be obtained in \cite{rowlett2020decisions,pedro2020conditions}. 

\section{Conclusions and prospects}

In summary, we have provided a brief review of eco-evolutionary dynamics with strategy-dependent environmental feedbacks, including theoretical frameworks from two-player games to public goods game, the open problem of steering eco-evolutionary games with external control laws and application cases of such framework in different topics.         

As stressed, the feedback-evolving game is certainly a promising filed to investigate, as it characterizes the intricate interplay of cooperation and environment in different scales of coupled dynamical systems. In particular, we believe three research directions will be of wide concern for interdisciplinary physics: (1) Theoretically, current works mainly focus on classic two-player games or public goods game with linear environmental feedbacks in well-mixed populations. However, a more complete picture of the increasing complexity arising from both the nonlinearity of coupled evolutions and the structured interactions remains unclear. For instance, will the strategic and environmental outcomes be qualitatively different in nonlinear PGG with nonlinear feedbacks? What role will the complex network structures play in mitigating social dilemma? The eco-evolutionary dynamics in meta-population with heterogeneous environments is also interesting for future exploration. Besides, many other game types are also worth considering, such as the rock-scissors-paper game, the ultimatum game and the stochastic games \cite{liao2019rock, hilbe2018evolution}.  (2) The control mechanism that can steer the system to a desired state is of vital significance for solving many social dilemmas. Nevertheless, as highlighted before, this is still an open problem which demands further explorations. Perhaps a potential way is to combine coevolutionary game theory with control theory.  (3) Finally, the more specific modelings that can be directly applied into diverse fields are greatly needed, ranging from biology (feedback loops in genomics \cite{becks2012functional}), public health (vaccine \cite{bauch2004vaccination,bauch2003group,fu2011imitation,chen2019imperfect}, antibiotic use \cite{chen2018social}, co-evolutions of behavior-disease interactions \cite{rowlett2020decisions,pedro2020conditions,fu2017dueling}), social science (coevolution of behavior-belief systems \cite{fu2008coevolutionary,wang2020effect}, sustaining cooperation in a polarizing society \cite{wang2020public,liu2020homogeneity,fu2012evolution}, social dilemmas in online social networks \cite{fu2007social}), economics (how marketing environment changes, i.e., from bull to bear or from bear to bull, reshape the contracts game between firms and employees \cite{szolnoki2019seasonal}), to urgent global issues (climate change \cite{samset2020delayed}, overexploitation of natural resources \cite{cohen1995population}). The list could undoubtedly be much longer and we just name a few. Further, the integration of model and real data would be expected to provide more profound insights. 

The ongoing SARS-CoV-2 pandemic and the uncertainty due to this pandemic remind us that we are actually living in a world of constant change with many unpredictable accidents. The eco-evolutionary game dynamics indeed provides a valuable tool for understanding and predicting population behaviors in such a changing world. While so many big challenges still lie ahead, we believe sustained large-scale cooperations is necessary for coping with these imminent challenges \cite{hauser2014cooperating}. 

\acknowledgments
This work is supported by Program of National Natural Science Foundation of China Grant No. 11871004, 11922102, and National Key Research and Development Program of China Grant No. 2018AAA0101100. X.W. gratefully acknowledges generous support by the China Scholarship Council. F.F. is supported by a Junior Faculty Fellowship awarded by the Dean of the Faculty at Dartmouth and also by the Bill \& Melinda Gates Foundation (award no. OPP1217336), the NIH COBRE Program (grant no. 1P20GM130454), the Neukom CompX Faculty Grant, the Dartmouth Faculty Startup Fund, and the Walter \& Constance Burke Research Initiation Award.


\clearpage 

\end{document}